\documentclass[%
 reprint,
nofootinbib,
 amsmath,amssymb,
 aps,
]{revtex4-1}

\usepackage{graphicx}

\usepackage{dcolumn}% Align table columns on decimal point
\usepackage{bm}% bold math

\usepackage{multirow}

\usepackage{here}

\usepackage{ulem}
\usepackage{color}

\begin{document}

\title{Suppression of decay widths in singly heavy baryons induced by the $U_A (1)$ anomaly}

\author{Yohei Kawakami}
\email{kawakami@hken.phys.nagoya-u.ac.jp}
\affiliation{Department of Physics, Nagoya University, Nagoya, 464-8602, Japan}
\author{Masayasu Harada}
\email{harada@hken.phys.nagoya-u.ac.jp}
\affiliation{Department of Physics, Nagoya University, Nagoya, 464-8602, Japan}
\author{Makoto Oka}
\email{oka@post.j-parc.jp}
\affiliation{Advanced Science Research Center, Japan Atomic Energy Agency (JAEA), Tokai 319-1195, Japan}
\affiliation{Nishina Center for Accelerator-Based Science, RIKEN, Wako 351-0198, Japan}
\author{Kei Suzuki}
\email{k.suzuki.2010@th.phys.titech.ac.jp}
\affiliation{Advanced Science Research Center, Japan Atomic Energy Agency (JAEA), Tokai 319-1195, Japan}

\date{\today}

\begin{abstract}
We study strong and radiative decays of excited singly heavy baryons (SHBs) using an effective chiral Lagrangian based on the diquark picture proposed in Ref.~\cite{Harada:2019udr}.
The effective Lagrangian contains a $U_A (1)$ anomaly term, which induces an inverse mass ordering between strange and non-strange SHBs with spin-parity $1/2^-$. We find that the effect of the $U_A (1)$ anomaly combined with flavor-symmetry breaking modifies the Goldberger-Treiman relation for the mass difference between the ground state $\Lambda_Q (1/2^+)$ and its chiral partner $\Lambda_Q (1/2^-)$, and $\Lambda_Q (1/2^-) \Lambda_Q (1/2^+) \eta$ coupling, which results in suppression of the decay width of $\Lambda_Q (1/2^-) \to \Lambda_Q (1/2^+) \eta$. 
We also investigate the other various decays such as $\Lambda_Q (1/2^-) \to \Sigma_Q (1/2^+, \, 3/2^+) \pi \pi$, $\Lambda_Q (1/2^-) \to \Sigma_Q (1/2^+) \pi$, $\Lambda_Q (1/2^-) \to \Sigma_Q (1/2^+, \, 3/2^+) \gamma$, and $\Lambda_Q (1/2^-) \to \Lambda_Q (1/2^+) \pi^0$ for wide range of mass of $\Lambda_Q (1/2^-)$.
\end{abstract}

\maketitle

\section{\label{sec:level1}Introduction}
Spontaneous chiral symmetry breaking and the $U_A (1)$ anomaly are the essential properties of quantum chromodynamics (QCD). Since colored quarks and gluons are not directly observed at the low-energy scale in QCD, verification of these properties in hadronic phenomena provides precious clues to understand the symmetry properties of QCD. Chiral partner structure of hadron spectra and the heavy $\eta^\prime$ mass spectrum are known as the examples of such phenomena.

It is also important to interpret hadronic phenomena based on colored constitutions such as diquarks.
The diquark is the simplest colored cluster, that is known to play important roles in structures of baryons and exotic multi-quark hadrons, and color superconducting phase.

Singly heavy baryons (SHBs) are considered and studied as the bound states of a diqaurk and a heavy quark ($c$ or $b$ quark).
Recently, diquarks made of light quarks are studied from the chiral-symmetry viewpoints and a chiral effective theory for scalar/pseudo-scalar diquarks was proposed~\cite{Harada:2019udr}.
The proposed Lagrangian contains a term representing $U_A (1)$ anomaly effect.
It is found that the term induces the inverse mass ordering between strange and non-strange SHBs with spin parity $1/2^-$.

In this paper, we focus on investigation into decay widths of SHBs with spin-parity $1/2^-$ based on the model given in Ref.~\cite{Harada:2019udr}, and we find that the effect of $U_A (1)$ anomaly combined with the flavor-symmetry breaking modifies the Goldberger-Treiman (GT) relation for the mass difference between $\Lambda_Q (1/2^-)$ and $\Lambda_Q (1/2^+)$, and $\Lambda_Q (1/2^-) \Lambda_Q (1/2^+) \eta$ coupling. This modification induces suppression of the decay width of $\Lambda_Q (1/2^-) \to \Lambda_Q (1/2^+) \eta$, when the mass of $\Lambda_Q (1/2^-)$ is above the threshold of $\Lambda_Q (1/2^+) \eta$. We also study various other decay modes of $\Lambda_Q (1/2^-)$   for the mass region below the threshold such as $\Lambda_Q (1/2^-) \to \Sigma_Q (1/2^+, \, 3/2^+) \pi \pi$, $\Lambda_Q (1/2^-) \to \Sigma_Q (1/2^+) \pi$, $\Lambda_Q (1/2^-) \to \Sigma_Q (1/2^+, \, 3/2^+) \gamma$, and $\Lambda_Q (1/2^-) \to \Lambda_Q (1/2^+) \pi^0$.
We finally mention decays of $\Xi_Q (1/2^-)$ based on the model.

This paper is organized as follows:
We show the masses and GT relations obtained in the model in Sec.~\ref{sec:masses}.
Section~\ref{sec:eta} is 
devoted to study the effect of the $U_A (1)$ anomaly to $\Lambda_Q (1/2^-) \to \Lambda_Q (1/2^+) \eta$ decay.
We study the other decays of $\Lambda_Q (1/2^-)$ in Sec.~\ref{sec:other}.
Finally, we give a summary and discussions in Sec.~\ref{sec:summary}.

\section{Masses and Goldberger-Treiman relations}
\label{sec:masses}

In Ref.~\cite{Harada:2019udr}, a chiral effective Lagrangian of scalar and pseudo-scalar diquarks based on chiral $SU(3)_R \times SU(3)_L$ symmetry is proposed.
Each diquark with a heavy quark $Q \, (c/b)$  makes an SHB as a bound state which belongs to the flavor $\bar{3}$ representation ($\Lambda_{c/b}$ or $\Xi_{c/b}$).
In this paper, we express those SHBs by linear representations: $S_{R, i}$ ($i = 1, \, 2, \, 3$) belongs to $(\bar{3}, \, 1)$ representation under $SU(3)_R \times SU(3)_L$ symmetry, $S_{L, i}$ to $(1, \, \bar{3})$.
The effective Lagrangian of the SHBs in the chiral limit is given as
\begin{align}
	\mathcal{L} = & \ \bar{S}_{R, i} (i v^\mu \partial_\mu) S_{R, i} + \bar{S}_{L, i} (i v^\mu \partial_\mu) S_{L, i} \notag \\
	& -M_{B 0} \left( \bar{S}_{R, i} S_{R, i} +\bar{S}_{L, i} S_{L, i}\right) \notag \\
	& -\frac{M_{B 1}}{f} \left( \bar{S}_{R, i} \Sigma^T_{ij} S_{L, j} +\bar{S}_{L, i} \Sigma^{T \dagger}_{ij} S_{R, j}\right) \notag \\
	& -\frac{M_{B2}}{2 f^2} \epsilon_{ijk} \epsilon_{lmn} \left( \bar{S}_{L, k} \Sigma^T_{li} \Sigma^T_{mj} S_{R, n} +\bar{S}_{R, k} \Sigma^{T \dagger}_{li} \Sigma^{T \dagger}_{mj} S_{L, n}\right), \label{eq:lag}
\end{align}
where $v^\mu$ is a velocity of SHBs, $M_{B0}$, $M_{B1}$, and $M_{B2}$ are model parameters, $f=92.4$ MeV is the pion decay constant,
the indices $i, \, j, \, k, \, l, \, m, \, n = 1, \, 2, \, 3$ are for either $SU(3)_R$ or $SU(3)_L$, and summations over repeated indices are understood.
$\Sigma_{ij}$ denotes the effective field for light scalar and pseudo-scalar mesons belonging to the chiral $(\bar{3}, \, 3)$ representation.
These fields transform as
\begin{align}
	\Sigma_{ij} &\to U_{L, ik} \Sigma_{kl} U_{R, lj}^\dagger, \\
	S_{R, i} &\to U_{R, ji}^\dagger S_{R, j}, \quad S_{L, i} \to U_{L, ji}^\dagger S_{L, j}.
\end{align}
The Lagrangian is invariant under these chiral transformations.
In addition, the kinetic, $M_{B0}$-, and $M_{B2}$-terms are also invariant under the following $U_A (1)$ transformations:
\begin{align}
	\Sigma_{ij} &\to e^{-2i\theta}\Sigma_{ij}, \\
	S_{R, i} &\to e^{2i\theta} S_{R, i}, \quad S_{L, i} \to e^{-2i\theta} S_{L, i}.
\end{align}
In contrast, the $M_{B1}$-term is not invariant under these transformations, reflecting the $U_A (1)$ anomaly.

The chiral symmetry is spontaneously broken by the vacuum expectation values of $\Sigma_{ij}$ field as $\langle \Sigma_{ij} \rangle = f \delta_{ij}$.
Then, the $M_{B1}$- and $M_{B2}$-terms give contributions to the mass splitting between parity eigenstates of SHBs defined as
\begin{equation}
	S_i = \frac{1}{\sqrt{2}} (S_{R, i} -S_{L,i}) = 
  \left\{
    \begin{array}{ll}
      \Xi_Q (1/2^+) & (i = 1, \, 2) \\
      \Lambda_Q (1/2^+) & (i = 3),
    \end{array}
  \right. \label{eq:positive}
\end{equation}
\begin{equation}
	P_i = \frac{1}{\sqrt{2}} (S_{R, i} +S_{L,i}) = 
  \left\{
    \begin{array}{ll}
      \Xi_Q (1/2^-) & (i = 1, \, 2) \\
      \Lambda_Q (1/2^-) & (i = 3).
    \end{array}
  \right. \label{eq:negative}
\end{equation}

In this paper, we follow the prescription adopted in Ref.~\cite{Harada:2019udr}, in which the explicit breaking of flavor symmetry is introduced by the replacement,
\begin{equation}
	\Sigma \to \tilde{\Sigma} \equiv \Sigma +{\rm{diag}}\left \{0, \, 0, \, (A-1)f \right \},
\end{equation}
with $A \sim 5/3$ being the parameter of flavor breaking. The vacuum expectation value of $\tilde{\Sigma}$ is given by
\begin{equation}
	\langle \tilde{\Sigma} \rangle = f {\rm{diag}} (1, 1, A).
\end{equation}
Then, the masses of the SHBs are expressed as
\begin{align}
	M_{1, 2}^\pm & = M_{B0} \mp \left( M_{B1} +A M_{B2}\right) \label{eq:M12}, \\
	M_3^\pm & = M_{B0} \mp \left( A M_{B1} +M_{B2}\right) \label{eq:M3},
\end{align}
where $M_{1, 2}^\pm$ denote the masses of $\Xi_Q (1/2^+)$ and $\Xi_Q (1/2^-)$, and  $M_3^\pm$ the masses of $\Lambda_Q (1/2^+)$ and $\Lambda_Q (1/2^-)$.
From Eqs.~(\ref{eq:M12}) and (\ref{eq:M3}), we obtain mass differences between chiral partners as
\begin{align}
	\Delta M_{1, 2} & = 2 \left(M_{B1} +A M_{B2} \right) \label{eq:dM12}, \\
	\Delta M_3 & = 2 \left(A M_{B1} +M_{B2} \right) \label{eq:dM3}.
\end{align}
We require $M_{1, 2}^+ > M_3^+$ consistently with the experimental values of the masses of the ground-state SHBs. The inverse mass ordering between strange and non-strange SHBs proposed in Ref.~\cite{Harada:2019udr} indicates $M_{1, 2}^- < M_3^-$, then we obtain a relation $\Delta M_{1, 2} < \Delta M_3$. For this relation, the effect of anomaly plays a crucial role. When we ignore $M_{B1}$ in Eqs.~(\ref{eq:M12})-(\ref{eq:dM3}),  the realistic mass ordering of SHBs ($M_{1, 2}^+ < M_{1, 2}^-$ and $M_3^+ < M_3^-$) results in $M_{B2} > 0$.
Then, $M_{B2} > 0$ together with $A > 1$ leads to $\Delta M_{1, 2} > \Delta M_3$. 

Next, we study the couplings among chiral partners and a pseudo Nambu-Goldstone (pNG) boson.
For this purpose, we introduce
light scalar mesons $\sigma_{ij}$ and pseudo-scalar mesons $\pi_{ij}$ as
\begin{equation}
	\tilde{\Sigma}_{ij} = \langle \tilde{\Sigma}_{ij} \rangle + \sigma_{ij} +i \pi_{ij},
\end{equation}
with
\begin{equation}
	\pi_{ij} = \sqrt{2} \begin{pmatrix} \frac{\pi^0}{\sqrt{2}} +\frac{\eta_8}{\sqrt{6}} +\frac{\eta_1}{\sqrt{3}} & \pi^+ & K^+ \\
	\pi^- & -\frac{\pi^0}{\sqrt{2}} +\frac{\eta_8}{\sqrt{6}} +\frac{\eta_1}{\sqrt{3}} & K^0 \\
	K^- & \bar{K}^0 & -\frac{2\eta_8}{\sqrt{6}} +\frac{\eta_1}{\sqrt{3}} \end{pmatrix}. \label{eq:pi_matrix}
\end{equation}

The $M_{B1}$- and $M_{B2}$-terms of the Lagrangian (\ref{eq:lag}) provide interactions of SHBs with light mesons.
In the chiral limit ($A = 1$), we obtain the relation between the coupling constant for the interaction of SHBs with a pNG boson $g_{\pi S P}$ and the mass difference of SHBs $\Delta M$ as
\begin{equation}
	g_{\pi S P} = \frac{M_{B1} +M_{B2}}{f} =\frac{\Delta M}{2f}. \label{eq:gtcl}
\end{equation}
This is often called the extended GT relation.
We focus on studying the coupling of $\Lambda_Q (1/2^-) \Lambda_Q (1/2^+) \eta$ in this section,
and we use $g$ for the coupling constant below.
When the flavor-symmetry breaking is included by $A > 1$, the coupling constant is obtained as
\begin{equation}
	g = \frac{M_{B1} +M_{B2}}{f} = \frac{\Delta M_{1,2} +\Delta M_3}{2 f (A +1)} \label{eq:gt3}.
\end{equation}
From inverse mass ordering, we obtain $\Delta M_{1, 2} < \Delta M_3$ as we showed above.
Then, we see that
\begin{equation}
	g = \frac{\Delta M_3}{2 f} \frac{\frac{\Delta M_{1,2}}{\Delta M_3} +1}{A +1} < \frac{\Delta M_3}{2 f}, \label{eq:gtmod}
\end{equation}
which indicates that the value of the coupling constant is smaller than the one expected from the GT relation. 
In order to see that this is caused by the effect of anomaly,
we drop $M_{B1}$ in Eqs.~(\ref{eq:dM3}) and (\ref{eq:gt3}). Then, we obtain the coupling constant as
\begin{equation}
	\bar{g} = \frac{M_{B2}}{f} = \frac{\Delta M_3}{2 f} \label{eq:no-anomaly},
\end{equation}
which is the one expected from the GT relation in Eq.~(\ref{eq:gtcl}).
Therefore, we conclude that the $U_A (1)$ anomaly suppresses the value of coupling constant. This suppression is expected to be seen in the decay width of $\Lambda_Q (1/2^-) \to \Lambda_Q (1/2^+) \eta$.

\section{Effect of anomaly to $\Lambda_Q (1/2^-) \to \Lambda_Q (1/2^+) \eta$ decay}
\label{sec:eta}
In this section, we numerically study how the effect of anomaly suppresses the decay of $\Lambda_Q (1/2^-) \to \Lambda_Q (1/2^+) \eta$.
Interaction among $\Lambda_Q (1/2^-)$, $\Lambda_Q (1/2^+)$, and $\eta_8$ or $\eta_1$ is obtained from the Lagrangian (\ref{eq:lag}) as
\begin{align}
	i\frac{2}{\sqrt{3}} &\left \{ \frac{M_{B1}}{f} \left(\eta_8 -\frac{1}{\sqrt{2}} \eta_1 \right) +\frac{M_{B2}}{f} \left(\eta_8 +\sqrt{2} \eta_1 \right) \right\} \notag \\
	&\cdot \bar{\Lambda}_Q (1/2^+) \Lambda_Q (1/2^-), \label{eq:coupling}
\end{align}
where $\eta_8$ is a member of the octet of $SU (3)$ flavor symmetry, and $\eta_1$ belongs to the flavor singlet. The realistic $\eta$ is known as a mixing state of $\eta_8$ and $\eta_1$.
As shown in the previous section, 
although $\eta_8$ is a pNG boson associated with the chiral $SU(3)_R \times SU(3)_L$ symmetry breaking, its coupling constant to $\Lambda_c (1/2^-)$ and $\Lambda_c (1/2^+)$ given in Eq.~(\ref{eq:gt3}) is smaller than the naive expectation of the GT relation in Eq.~(\ref{eq:gtcl}). On the other hand, the coupling constant of $\eta_1$ is read from Lagrangian (\ref{eq:lag}) as
\begin{equation}
	g_{\eta_1} = \frac{M_{B1} -2M_{B2}}{f}.
\end{equation}
This is also different from the GT relation, since $\eta_1$ is no longer a pNG boson when the effect of $U_A (1)$ anomaly is included. 
In the following, we first see that the coupling constant in Eq.~(\ref{eq:gt3}) is suppressed from the one in Eq.~(\ref{eq:no-anomaly}) due to the effect of anomaly as shown in Eq.~(\ref{eq:gtmod})
by regarding $\eta_8$ as $\eta$ with neglecting small mixing between $\eta$ and $\eta^\prime$.
Effect of the $\eta$-$\eta^\prime$ mixing is introduced for comparison with the realistic decay width afterward.
The values of $M_{B1}$ and $M_{B2}$ in Eq.~(\ref{eq:gt3}) are determined from the masses of $\Lambda_c (1/2^-)$, $\Lambda_c (1/2^+)$, and $\Xi_c (1/2^+)$. We use the experimental values of masses of $\Lambda_c (1/2^+)$ and $\Xi_c (1/2^+)$ as $M\left(\Lambda_c (1/2^+)\right) = 2286.46$ MeV  and $M\left(\Xi_c (1/2^+)\right) = 2469.42$ MeV. \footnote{When we calculate the coupling constant in Eq.~(\ref{eq:no-anomaly}), we do not use the mass of $\Xi_c (1/2^+)$ as an input.}
Since the mass of $\Lambda_c (1/2^-)$ is not determined in experiment, we calculate the decay width for wide range of the mass of $\Lambda_c (1/2^-)$.
We note that, although $\Lambda_c (2595)$ carries $J^P = 1/2^-$, it is not $\Lambda_c (1/2^-)$ here since $\Lambda_c (2595)$ makes a heavy-quark spin doublet with $\Lambda_c (2625)$ having $3/2^-$.

\begin{figure}[htbp]
  \begin{center}
   \includegraphics[bb = 0 0 432 288,width=70mm]{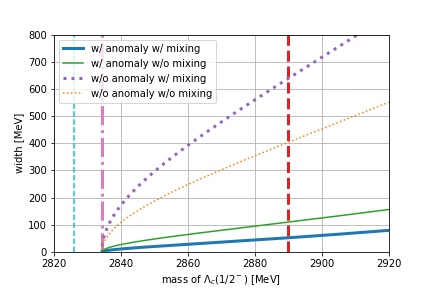}
  \end{center}
  \caption{Dependence of $\Lambda_c (1/2^-) \to \Lambda_c (1/2^+) \eta$ width on the mass of  $\Lambda_c (1/2^-)$.
Predictions without the $\eta$-$\eta^\prime$ mixing are shown by the thin-solid green (with anomaly) and thin-dotted orange (without anomaly) curves,
and those with the $\eta$-$\eta^\prime$ mixing are by the thick-solid blue and thick-dotted purple curves.
Two predictions of the mass of $\Lambda_c (1/2^-)$ are shown by the vertical thick-dashed red \cite{Yoshida:2015tia} and thin-dashed cyan \cite{Harada:2019udr} lines.
The thick-dash-dotted magenta line shows the threshold of $\Lambda_c (1/2^+) \eta$ ($\sim 2834$ MeV).}
  \label{fig:charm_eta}
\end{figure}

\begin{figure}[htbp]
  \begin{center}
   \includegraphics[bb = 0 0 432 288,width=70mm]{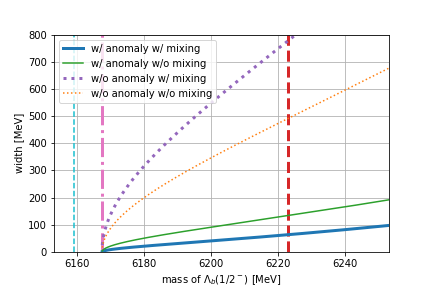}
  \end{center}
  \caption{Dependence of $\Lambda_b (1/2^-) \to \Lambda_b (1/2^+) \eta$ width on the mass of  $\Lambda_b (1/2^-)$.
Curves in this figure respectively correspond to those in Fig.~\ref{fig:charm_eta}.
See the main text for the vertical lines.}
  \label{fig:bottom_eta}
\end{figure}

We show the decay width of $\Lambda_c (1/2^-) \to \Lambda_c (1/2^+) \eta$ in Fig.~\ref{fig:charm_eta}.
The thin-solid green and thin-dotted orange curves are plotted without $\eta$-$\eta^\prime$ mixing.
The thin-solid green curve is drawn by using the coupling constant in Eq.~(\ref{eq:gt3}) which includes the effect of anomaly, while the thin-dotted orange curve is by the one in Eq.~(\ref{eq:no-anomaly}) without anomaly.
One can easily see that the thin-solid green curve is much suppressed compared with the thin-dotted orange curve.

Next, let us include the effect of  $\eta$-$\eta^\prime$ mixing.
Introducing the $\eta$-$\eta^\prime$ mixing matrix as~\cite{Tanabashi:2018oca},
\begin{equation}
	\begin{pmatrix} \eta \\ \eta^\prime \end{pmatrix} = \begin{pmatrix} \cos \theta_P & -\sin \theta_P \\ \sin \theta_P & \cos \theta_P \end{pmatrix} \begin{pmatrix} \eta_8 \\ \eta_1 \end{pmatrix},
\end{equation}
$\Lambda_Q (1/2^-) \Lambda_Q (1/2^+) \eta$ interaction is obtained from Eq.~(\ref{eq:coupling}) as
\begin{align}
	i\frac{2}{\sqrt{3}} g_{\rm{phys}} \bar{\Lambda}_Q (1/2^+) \Lambda_Q (1/2^-) \eta,
\end{align}
where
\begin{equation}
	g_{\rm{phys}} = \frac{\xi_1 M_{B1} +\xi_2 M_{B2}}{f}, \label{eq:mixed}
\end{equation}
with $\xi_1 = \cos \theta_P +\sin \theta_P / \sqrt{2}$ and $ \xi_2 = \cos \theta_P -\sqrt{2} \sin \theta_P$. Here, we use $\theta_P = -11.3^\circ$ listed in Ref.~\cite{Tanabashi:2018oca}. 
In Fig.~\ref{fig:charm_eta},
the thick-solid blue curve shows the width of $\Lambda_c (1/2^-) \to \Lambda_c (1/2^+) \eta$ decay calculated by using the coupling constant in Eq.~(\ref{eq:mixed}). Comparing with the thin-solid green curve, we can see that the effect of $\eta$-$\eta^\prime$ mixing suppresses the decay width by about $50$ \%.
When the effect of anomaly is dropped with taking $M_{B1} = 0$, the coupling constant in Eq.~(\ref{eq:mixed}) is reduced to
\begin{equation}
	\bar{g}_{\rm{phys}} = \frac{\xi_2 M_{B2}}{f} = \xi_2 \bar{g}. \label{eq:no-anomaly mixed}
\end{equation}
Since $\xi_2 \sim 1.26$, the above relation implies that the width is enhanced by about $60$ \% as shown by the thick-dotted purple curve compared with the thin-dotted orange curve in Fig.~\ref{fig:charm_eta}.

At the last of this section, let us apply our analysis to the bottom sector.
The parameters $M_{B1}$ and $M_{B2}$ of the Lagrangian (\ref{eq:lag}) are uniquely determined when the mass of $\Lambda_c (1/2^-)$ on the horizontal axis in Fig.~\ref{fig:charm_eta} is fixed.
Here, we apply the Lagrangian (\ref{eq:lag}) to the bottom sector
and use the values of $M_{B1}$ and $M_{B2}$ determined above to evaluate the mass difference between $\Lambda_b (1/2^+)$ and $\Lambda_b (1/2^-)$, and the decay width of $\Lambda_b (1/2^-) \to \Lambda_b (1/2^+) \eta$.
The mass difference between $\Lambda_b (1/2^+)$ and $\Lambda_b (1/2^-)$ is equal to that between $\Lambda_c (1/2^+)$ and $\Lambda_c (1/2^-)$.
We show the resultant decay width in Fig.~\ref{fig:bottom_eta}.
In this figure, the vertical thick-dashed red line shows $M(\Lambda_b, \, 1/2^-) = 6223$ MeV determined by using $M(\Lambda_c, \, 1/2^-) = 2890$ MeV \cite{Yoshida:2015tia} as an input,
and the vertical thin-dashed cyan line is for $M(\Lambda_b, \, 1/2^-) = 6159$ MeV by $M(\Lambda_c, \, 1/2^-) = 2826$ MeV \cite{Harada:2019udr}.
The vertical thick-dash-dotted magenta line shows the threshold of $\Lambda_b (1/2^+) \eta$.
We can see that the effect of anomaly suppresses the decay widths as in the charm sector.
Similarly, the effect of $\eta$-$\eta^\prime$ mixing enhances the decay width in case without anomaly,
and suppresses that with anomaly.
We also observe that all the decay widths of $\Lambda_b (1/2^-)$ in Fig.~\ref{fig:bottom_eta} are enhanced compared with those of $\Lambda_c (1/2^-)$ in Fig.~\ref{fig:charm_eta}.
This is caused by the kinematical factor, although the relevant coupling constants are equal to each other.

\section{Other decays of $\Lambda_Q (1/2^-)$}
\label{sec:other}
The mass of $\Lambda_Q (1/2^-)$ ($Q = c, \, b$) is not yet experimentally determined.
When it is larger than the threshold of $\Lambda_Q (1/2^-) \eta$, $\Lambda_Q (1/2^-) \to \Lambda_Q (1/2^+) \eta$ decay is expected to be dominant.
On the other hand, if $\Lambda_Q (1/2^-)$ locates below the threshold
as predicted in Refs.~\cite{Harada:2019udr, Kim:2020imk},
other decay modes become relevant.
Then, we investigate decays of
$\Lambda_Q (1/2^-) \to \Sigma_Q^{(\ast)} \pi \pi$, $\Lambda_Q (1/2^-) \to \Sigma_Q^{(\ast)} \gamma,$ \, $\Lambda_Q (1/2^-) \to \Sigma_Q \pi$ and $\Lambda_Q (1/2^-) \to \Lambda_Q (1/2^+) \pi^0$ ($\Sigma_Q$ and $\Sigma_Q^\ast$ denote $\Sigma_Q (1/2^+)$ and $\Sigma_Q (3/2^+)$ respectively).

We expect that $\Lambda_Q (1/2^-) \to \Sigma_Q^{(\ast)} \pi \pi$ decay becomes large when the mass of $\Lambda_Q (1/2^-)$ is far above the threshold of $\Lambda_Q (1/2^-) \to \Sigma_Q \pi \pi$,
and that the radiative decay is suppressed.
Although $\Lambda_Q (1/2^-) \to \Sigma_Q \pi$ decay violates the heavy-quark symmetry,
it can be dominant near the threshold of $\Lambda_Q (1/2^-) \to \Sigma_Q \pi \pi$ especially in the charm sector.
$\Lambda_Q (1/2^-) \to \Lambda_Q (1/2^+) \pi^0$ will be strongly suppressed since it breaks isospin symmetry.

Typical forms of interaction Lagrangians are given in Appendix \ref{sec:33}.
The coupling constant of $\Lambda_Q (1/2^-) \to \Sigma_Q^{(\ast)} \pi \pi$ is estimated as $k = 1$ by the $\rho$ meson dominance and the coupling universality~\cite{Sakurai:1969, Bando:1987br, Harada:2003jx}, but so far we cannot precisely determine its value using the known experimental data of SHBs.
Furthermore, the coupling constants of $\Lambda_Q (1/2^-) \to \Sigma_Q \pi$ and $\Lambda_Q (1/2^-) \to \Sigma_Q^{(\ast)} \gamma$ are unknown.
We therefore leave $k$, $\kappa$, and $r$ as free parameters.

\begin{figure}[H]
  \begin{center}
   \includegraphics[bb= 0 0 432 288, width=70mm]{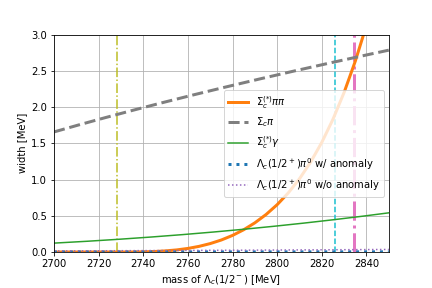}
  \end{center}
  \caption{
Dependence of various decay widths of $\Lambda_c (1/2^-)$ on the mass of $\Lambda_c (1/2^-)$.
The thick-solid orange, thick-dashed gray, and thin-solid green curves respectively show $\Gamma_{\Sigma_c^{(\ast)} \pi \pi}/k^2$, $\Gamma_{\Sigma_c \pi}/\kappa^2$, and $\Gamma_{\Sigma_c^{(\ast)} \gamma}/r^2$. The thick-dotted blue/thin-dotted purple curve shows the width of $\Lambda_c (1/2^-) \to \Lambda_c (1/2^+) \pi^0$ with the coupling constant in Eq.~(\ref{eq:pi0}) (with anomaly)/Eq.~(\ref{eq:pi0 no-anomaly}) (without anomaly). The vertical thick-dash-dotted magenta and thin-dashed cyan lines correspond to those in Fig.~{\ref{fig:charm_eta}}. The vertical thin-dash-dotted olive line shows the threshold of $\Sigma_c \pi \pi$.
}
  \label{fig:charm_various}
\end{figure}

\begin{figure}[H]
  \begin{center}
   \includegraphics[bb=0 0 432 288 , width=70mm]{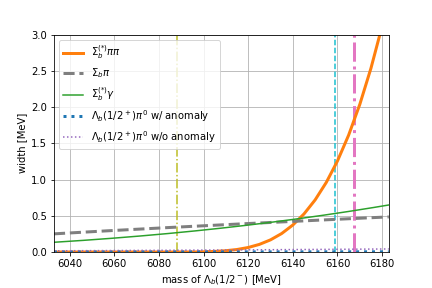}
  \end{center}
  \caption{
Dependence of various decay widths of $\Lambda_b (1/2^-)$ on the mass of $\Lambda_b (1/2^-)$.
The curves correspond to those in Fig.~\ref{fig:charm_various}. The vertical thick-dash-dotted magenta and thin-dashed cyan lines correspond to those in Fig.~\ref{fig:bottom_eta}. The vertical thin-dash-dotted olive line shows the threshold of $\Sigma_b \pi \pi$.
}
  \label{fig:bottom_various}
\end{figure}

We show the dependence of estimated widths on the mass of $\Lambda_c (1/2^-)$ in Fig.~\ref{fig:charm_various}.
We also show the estimated widths of $\Lambda_b (1/2^-)$ in Fig.~\ref{fig:bottom_various}.
We note that these figures show the decay widths of $\Lambda_Q (1/2^-) \to \Sigma_Q^{(\ast)} \pi \pi$, $\Lambda_Q (1/2^-) \to \Sigma_Q \pi$, and $\Lambda_Q (1/2^-) \to \Sigma_Q^{(\ast)} \gamma$ divided by the unknown constants $k^2$, $\kappa^2$, and $r^2$, respectively.
On the other hand, the $\Lambda_Q (1/2^-) \to \Lambda_Q (1/2^+) \pi^0$ mode is completely determined by the chiral property in the present analysis, so that decay width itself is plotted.

We see that decay widths of $\Lambda_Q (1/2^-) \to \Sigma_Q^{(\ast)} \pi \pi$, $\Lambda_Q (1/2^-) \to \Sigma_Q^{(\ast)} \gamma$, and $\Lambda_Q (1/2^-) \to \Lambda_Q (1/2^+) \pi^0$ in the charm sector are almost the same as those in the bottom sector.
On the other hand, the decay width of $\Lambda_b (1/2^-) \to \Sigma_b \pi$ is much suppressed compared with that of $\Lambda_c (1/2^-) \to \Sigma_c \pi$ since the heavy-quark symmetry is well satisfied in the bottom sector.

Figure~\ref{fig:charm_various} shows that, below the threshold of $\Lambda_c (1/2^+) \eta$ indicated by the vertical thick-dash-dotted magenta line, the dominant decay mode is $\Lambda_c (1/2^-) \to \Sigma_c \pi$ shown by the thick-dashed gray curve which violates the heavy-quark symmetry.
Figure~\ref{fig:bottom_various} shows that, below the threshold, the dominant decay mode is $\Lambda_b (1/2^-) \to \Sigma_b^{(\ast)} \pi \pi$ indicated by the thick-solid orange curve.
When the mass of $\Lambda_b (1/2^-)$ is a little above the threshold of $\Lambda_b (1/2^-) \to \Sigma_b \pi \pi$ indicated by the vertical thin-dash-dotted olive line, $\Lambda_b (1/2^-) \to \Sigma_b^{(\ast)} \pi \pi$ decay is suppressed, and $\Lambda_b (1/2^-) \to \Sigma_b \pi$ and $\Lambda_b (1/2^-) \to \Sigma_b^{(\ast)} \gamma$ decays are dominant.
Since $\Lambda_b (1/2^-) \to \Sigma_b \pi$ decay is strongly suppressed by the heavy-quark symmetry in the bottom sector,
the width is comparable to that of the radiative decay.

$\Lambda_Q (1/2^-) \to \Lambda_Q (1/2^+) \pi^0$ decay is originated from $\pi^0$-$\eta$ mixing generated by the isospin violation. 
The coupling constant of $\Lambda_Q (1/2^-) \to \Lambda_Q (1/2^+) \pi^0$ is written as
\begin{equation}
	g_{\pi^0 \eta} = \Delta_{\pi^0 \eta} g_{\rm{phys}}, \label{eq:pi0}
\end{equation}
where $\Delta_{\pi^0 \eta}$ is a parameter of $\pi^0$-$\eta$ mixing estimated as $\Delta_{\pi^0 \eta} \sim -5.32 \times 10^{-3}$ in Ref.~\cite{Harada:2003kt} with a scheme given in Ref.~\cite{Harada:1995sj}. Similarly, we obtain
\begin{equation}
	\bar{g}_{\pi^0 \eta} = \Delta_{\pi^0 \eta} \bar{g}_{\rm{phys}}. \label{eq:pi0 no-anomaly}
\end{equation}
Predicted widths from Eq.~(\ref{eq:pi0}) are shown by thick-dotted blue curves in Figs.~\ref{fig:charm_various} and \ref{fig:bottom_various}, and those from Eq.~(\ref{eq:pi0 no-anomaly}) by thin-dotted purple curves for comparison.
We see that these decay widths are very small.


\section{A summary and discussions}
\label{sec:summary}

In this paper, we study strong and radiative decays of excited SHBs using a chiral effective Lagrangian based on the diquark picture proposed in Ref.~\cite{Harada:2019udr}. We show predictions of widths for typical choices of the mass of $\Lambda_Q (1/2^-)$ in Tables~\ref{tab:charm_width} and \ref{tab:bottom_width}. Our prediction on the width of $\Lambda_Q (1/2^-) \to \Lambda_Q (1/2^+) \eta$ is strongly suppressed by the effect of anomaly compared with the prediction without anomaly.
In general, the large width of the chiral partner state is an obstacle against its observation.
The suppression of the width may enable us to observe the state easily.
Tables \ref{tab:charm_width} and \ref{tab:bottom_width} also show predictions on the other decay modes which are useful for experimental observation of chiral partner when the mass of $\Lambda_Q (1/2^-)$ locates below the threshold of $\Lambda_Q (1/2^+) \eta$.

In the chiral partner structure using
$(3, \, 3)$ representations 
for the diquark with $J^P = 1^{\pm}$, the chiral partners to $\Sigma_Q(J^P = 1/2^+)$ and $\Sigma_Q^\ast(3/2^+)$ are $\Lambda_{Q1}(1/2^-)$ and $\Lambda_{Q1}^\ast(3/2^-)$, respectively as in Eq.~(\ref{def:S}) in Appendix~\ref{sec:33}.
In such a case, 
$\Lambda_Q (1/2^-) \to \Sigma_Q^{(\ast)} \gamma$ decays share a common coupling constant with $\Lambda_{Q1}^{(\ast)} \to \Lambda_Q (1/2^+) \gamma$ decays.
Once the chiral partner $\Lambda_{Q1}^{(\ast)}$ is identified with some physical particles such as in Refs.~\cite{Kawakami:2018olq, Kawakami:2019hpp}, we can check the chiral partner by seeing the radiative decays of those particles.

In Tables~\ref{tab:charm_width} and \ref{tab:bottom_width}, we can see that the decay width of $\Lambda_Q (1/2^-) \to \Lambda_Q (1/2^+) \pi^0$ is small.
Even though it might be difficult to observe this decay experimentally, it may give some informations on the $U_A (1)$ anomaly since its coupling constant is completely determined by the relation reflecting chiral symmetry shown in Eq.~(\ref{eq:pi0}) or (\ref{eq:pi0 no-anomaly}), and the width with the anomaly is strongly suppressed.
Then, we may check the effect of anomaly through this decay when the mass of $\Lambda_Q (1/2^-)$ locates below the threshold of $\Lambda_Q (1/2^+) \eta$.

\begin{widetext}
\begin{center}
\begin{table}[H]
\caption{Decay widths of $\Lambda_c (1/2^-)$ without and with the effect of anomaly. Units of masses and widths are in MeV.}
	\begin{center}
		\begin{tabular}{ccccc} \hline \hline
		 mass of $\Lambda_c (1/2^-)$ [MeV] & $2702$ \cite{Kim:2020imk} & $2759$ \cite{Kim:2020imk} & $2826$ \cite{Harada:2019udr} & $2890$ \cite{Yoshida:2015tia} \\ \hline
		 $\Lambda_c (1/2^-) \to \Lambda_c (1/2^+) \eta$ w/o anomaly & - & - & - & $639$ \\
		 $\Lambda_c (1/2^-) \to \Lambda_c (1/2^+) \eta$ w/ anomaly & - & - & - & $52.3$ \\
		 $\Lambda_c (1/2^-) \to \Sigma_c^{(\ast)} \pi \pi$ & - & $0.0400 k^2$ & $1.84 k^2$ & $13.0 k^2$ \\
		 $\Lambda_c (1/2^-) \to \Sigma_c \pi$ & $1.67 \kappa^2$ & $2.14 \kappa^2$ & $2.62 \kappa^2$ & $3.04 \kappa^2$ \\
		$\Lambda_c (1/2^-) \to \Sigma_c^{(\ast)} \gamma$ & $0.126 r^2$ & $0.243 r^2$ & $0.448 r^2$ & $0.718 r^2$ \\
		$\Lambda_c (1/2^-) \to \Lambda_c (1/2^+) \pi^0$ w/o anomaly & $0.0147$ & $0.0213$ & $0.0309$ & $0.0422$ \\
		$\Lambda_c (1/2^-) \to \Lambda_c (1/2^+) \pi^0$ w/ anomaly & $2.60 \times 10^{-4}$ & $7.87 \times 10^{-4}$ & $1.87 \times 10^{-3}$ & $3.46 \times 10^{-3}$ \\ \hline \hline
		\end{tabular}
	\end{center}
\label{tab:charm_width}
\end{table}

\begin{table}[H]
\caption{Decay widths of $\Lambda_b (1/2^-)$ without and with the effect of anomaly. Units of masses and widths are in MeV. ``$6159$ MeV'' is not listed in Ref.~\cite{Harada:2019udr}, but it is estimated with the same way of the prediction ``$2826$ MeV'' in Table~\ref{tab:charm_width}.}
	\begin{center}
		\begin{tabular}{cccccc} \hline \hline
		 mass of $\Lambda_b (1/2^-)$ [MeV] & $5999$ \cite{Kim:2020imk} & $6079$ \cite{Kim:2020imk} & $6159$ \cite{Harada:2019udr} & $6174$ \cite{Kim:2020imk} & $6207$ \cite{Kim:2020imk} \\ \hline
		 $\Lambda_b (1/2^-) \to \Lambda_b (1/2^+) \eta$ w/o anomaly & - & - & - & $223$ & $619$ \\
		$\Lambda_b (1/2^-) \to \Lambda_b (1/2^+) \eta$ w/ anomaly & - & - & - & $16.5$ & $52.9$ \\
		 $\Lambda_b (1/2^-) \to \Sigma_b^{(\ast)} \pi \pi$ & - & - & $1.16 k^2$ & $2.38 k^2$ & $7.95 k^2$ \\
		 $\Lambda_b (1/2^-) \to \Sigma_b \pi$ & $0.183 \kappa^2$ & $0.329 \kappa^2$ & $0.450 \kappa^2$ & $0.472 \kappa^2$ & $0.518 \kappa^2$ \\
		$\Lambda_b (1/2^-) \to \Sigma_b^{(\ast)} \gamma$ & $0.0795 r^2$ & $0.241 r^2$ & $0.531 r^2$ & $0.603 r^2$ & $0.781 r^2$ \\
		$\Lambda_b (1/2^-) \to \Lambda_b (1/2^+) \pi^0$ w/o anomaly & $0.0129$ & $0.0229$ & $0.0369$ & $0.0400$ & $0.0473$ \\
		$\Lambda_b (1/2^-) \to \Lambda_b (1/2^+) \pi^0$ w/ anomaly & $9.50 \times 10^{-5}$ & $7.41 \times 10^{-4}$ & $2.23 \times 10^{-3}$ & $2.62 \times 10^{-3}$ & $3.63 \times 10^{-3}$ \\ \hline \hline
		\end{tabular}
	\end{center}
\label{tab:bottom_width}
\end{table} 
\end{center}
\end{widetext}

The suppression of the decay width by the effect of anomaly can be also seen in the diquark level,
which we show in Appendix \ref{sec:diquark}.

\begin{figure}[H]
  \begin{center}
   \includegraphics[bb=0 0 432 288, width=70mm]{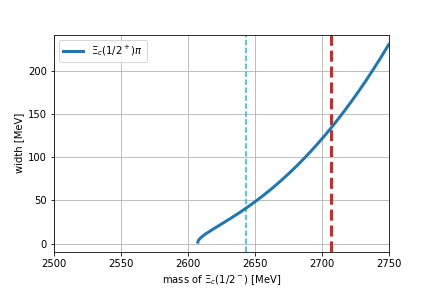}
  \end{center}
  \caption{Dependence of $\Xi_c (1/2^-) \to \Xi_c (1/2^+) \pi$ width on the mass of $\Xi_c (1/2^-)$. The vertical thick-dashed red line shows a prediction of mass of $\Xi_c (1/2^-)$ ($2707$ MeV) calculated from Eq.~(\ref{eq:Mxi}) using $M(\Lambda_c(1/2^-)) = 2890$ MeV predicted in Ref.~\cite{Yoshida:2015tia}, and the vertical thin-dashed cyan line shows a prediction ($2643$ MeV) using $M(\Lambda_c(1/2^-)) = 2826$ MeV predicted in Ref.~\cite{Harada:2019udr} .}
  \label{fig:Xi}
\end{figure}

We consider the mass and the decay width of $\Xi_c (1/2^-)$.
\footnote{
It should be noted here that the relevant chiral partner state,
$\Xi_c (1/2^-)$, is a singlet state for the heavy quark spin symmetry.
Therefore, the known $\Xi_c (2790) \, (1/2^-)$ may not be a candidate as it seems to form a heavy-quark spin doublet with $\Xi_c (2815) \, (3/2^-)$.
}
From Eqs.~(\ref{eq:M12}) and (\ref{eq:M3}), we obtain
\begin{align}
	& M (\Xi_c (1/2^-))  \notag \\ 
	&= M(\Lambda_c (1/2^+)) +M(\Lambda_c (1/2^-)) -M (\Xi_c (1/2^+)), \label{eq:Mxi}
\end{align}
which determines the mass of $\Xi_c (1/2^-)$, once the mass of $\Lambda_c (1/2^-)$ is fixed.
The dominant decay of $\Xi_c (1/2^-)$ is $\Xi_c (1/2^-) \to \Xi_c (1/2^+) \pi$, and the coupling constant is obtained from the Lagrangian (\ref{eq:lag}) as
\begin{equation}
	g_{\Xi_c} = \frac{M_{B1} +A M_{B2}}{f} = \frac{\Delta M_{1, 2}}{2 f}. \label{eq:gt12}
\end{equation}
This is the same as the extended GT relation in Eq.~(\ref{eq:gtcl}).
We plot the resultant relation between the mass of $\Xi_c (1/2^-)$ and the decay width of $\Xi_c (1/2^-) \to \Xi_c (1/2^+) \pi$ in Fig.~\ref{fig:Xi}. The vertical thick-dashed red and thin-dashed cyan lines correspond to those in Fig.~\ref{fig:charm_eta}: The thick-dashed red line shows $M(\Xi_c (1/2^-)) = 2707$ MeV obtained from $M (\Lambda(1/2^-)) = 2890$ MeV through Eq.~(\ref{eq:Mxi}); the thin-dashed cyan line shows $M(\Xi_c (1/2^-)) = 2643$ MeV from $M (\Lambda(1/2^-)) = 2826$ MeV.
The mass difference between $\Xi(1/2^+)$ and $\Xi(1/2^-)$ reflecting the inverse mass hierarchy
makes the phase space and the coupling constant small.

In the present analysis, we included only the leading order of the flavor symmetry breaking parameter, i.e., contributions are linear in $A$.
Furthermore, when the effect of heavy-quark symmetry violation is included,
$\Lambda_Q (1/2^-)$ can contain other diquark components which have different chiral properties~\cite{Dmitrasinovic:2020wye}.
We leave the study of higher orders of flavor symmetry breaking and mixing between different diquark components for future work.

\subsection*{Acknowledgments}
This work is supported in part by 
JSPS KAKENHI Grants No.~20K03927 (M. H.),
17K14277 (K. S.),
20K14476 (K. S.), and
19H05159 (M. O.).

\appendix

\section{Chiral $(3, \, 3)$ representation for SHBs and Lagrangian}
\label{sec:33}

To discuss various transitions of $\Lambda_Q (1/2^-)$ below the threshold of $\Lambda_Q (1/2^+) \eta$, we introduce heavy-quark spin doublet $\Sigma_Q^{(\ast)}$ as a member of chiral $(3, \, 3)$ field $S_{ij}^\mu$ (see e.g., Refs.~\cite{Kawakami:2018olq, Kawakami:2019hpp} for detailed discussion), which transforms as
\begin{equation}
	S_{ij}^\mu = \hat{B}^{6 \, \mu}_{ij} +\hat{B}^{\bar{3} \, \mu}_{ij} \to U_{R, ik} U_{L, jl} S_{kl}^\mu, \label{def:S}
\end{equation}
where $\hat{B}^{6 \, \mu}_{ij}$ is a doublet of $(1/2^+, \, 3/2^+)$ and $\hat{B}^{\bar{3} \, \mu}_{ij}$ is of $(1/2^-, \, 3/2^-)$. They are defined as matrix representations:
\begin{equation}
	\hat{B}^{6 \, \mu} = 
	\begin{pmatrix} \Sigma_Q^{I_3=1 \mu} & \frac{1}{\sqrt{2}} \Sigma_Q^{I_3=0 \mu} & \frac{1}{\sqrt{2}} \Xi_Q^{\prime I_3=\frac{1}{2} \mu} \\
	\frac{1}{\sqrt{2}} \Sigma_Q^{I_3=0 \mu} & \Sigma_Q^{I_3=-1 \mu} & \frac{1}{\sqrt{2}}\Xi_Q^{\prime I_3=-\frac{1}{2} \mu} \\
	\frac{1}{\sqrt{2}} \Xi_Q^{\prime I_3=\frac{1}{2} \mu} & \frac{1}{\sqrt{2}} \Xi_Q^{\prime I_3=-\frac{1}{2} \mu} & \Omega_Q^\mu
	\end{pmatrix},
\end{equation}
\begin{equation}
\hat{B}^{\bar{3} \, \mu} = \frac{1}{\sqrt{2}}
	\begin{pmatrix} 0 & \Lambda_{Q1}^\mu & \Xi_{Q1}^{I_3=\frac{1}{2} \mu} \\
	- \Lambda_{Q1}^\mu & 0 & \Xi_{Q1}^{I_3=-\frac{1}{2} \mu} \\
	- \Xi_{Q1}^{I_3=\frac{1}{2} \mu} & - \Xi_{Q1}^{I_3=-\frac{1}{2} \mu} & 0
	\end{pmatrix}.
\end{equation}
The Lagrangian of the interactions relevant for the present analysis is written as
\begin{align}
	& \mathcal{L}_{\rm{int}} = \notag \\
	& +\frac{k}{8f^3} \epsilon_{ijk} \left[ \Sigma_{il}^T \bar{S}_{lm}^\mu \left( \partial_\mu \Sigma^\dagger_{mn} \Sigma_{nj} -\Sigma_{mn}^\dagger \partial_\mu \Sigma_{nj} \right) S_{R, k} \right] \notag \\
	& +\frac{k}{8f^3} \epsilon_{ijk} \left[ \Sigma_{il}^{T \dagger} \bar{S}_{lm}^{T \mu} \left ( \partial_\mu \Sigma_{mn} \Sigma^\dagger_{nj} -\Sigma_{mn} \partial_\mu \Sigma^\dagger_{nj} \right )  S_{L, k} \right ] \notag \\
	& +\frac{\kappa f}{2 M_{\Lambda_Q}} \epsilon_{ijk}\epsilon_{\mu \nu \rho \sigma} \left( \bar{S}^\mu_{il} \Sigma^\dagger_{lj} v^\nu \sigma^{\rho \sigma} S_{L, k} -\bar{S}^{T\mu}_{il} \Sigma_{lj} v^\nu \sigma^{\rho \sigma} S_{R, k}\right) \notag \\
	& +\frac{r}{F^2} \epsilon_{ijk} \left( \bar{S}_{il}^\mu Q_{lm} \Sigma_{mj}^\dagger S_{L, k} +\bar{S}_{il}^{T \mu} Q_{lm} \Sigma_{mj} S_{R, k}\right) v^\nu F_{\mu \nu} \notag \\
	& + \rm{H.c.}, \label{eq:lag_int}
\end{align}
where $F_{\mu \nu}$ is a field strength of the photon and $\sigma_{\mu \nu}$ is defined as $\sigma_{\mu \nu} = \frac{i}{2} \left[\gamma_\mu, \, \gamma_\nu \right]$. The $\rho$ meson dominance and the coupling universality suggest $k = 1$~\cite{Sakurai:1969, Bando:1987br, Harada:2003jx}.
$M_{\Lambda_Q}$ denotes masses of the ground state $\Lambda_Q (1/2^+)$, and $F$ is a constant with dimension one. In this analysis, we take $F = 350$ MeV following Ref.~\cite{Cho:1994vg}.

\section{A decay width of diquark}
\label{sec:diquark}

In this appendix, we consider the decay width of diquark corresponding to $\Lambda_Q (1/2^-) \to \Lambda_Q (1/2^+) \eta$ decay. 

Let us first work in the chiral limit.
The Lagrangian of diquark is written as~\cite{Harada:2019udr},
\begin{align}
	\mathcal{L}_{qq} = & \mathcal{D}_{\mu} d_{R, i} \left( \mathcal{D}^\mu d_{R, i}\right)^\dagger +\mathcal{D}_{\mu} d_{L, i} \left( \mathcal{D}^\mu d_{L, i}\right)^\dagger \notag \\
	& -m_0^2 ( d_{R, i} d_{R, i}^\dagger +d_{L, i} d_{L, i}^\dagger ) \notag \\
	& -\frac{m_1^2}{f} ( d_{R, i} \Sigma_{ij}^\dagger d_{L, j}^\dagger +d_{L, i} \Sigma_{ij} d_{R, j}^\dagger) \notag \\
	& -\frac{m_2^2}{2 f^2} \epsilon_{ijk} \epsilon_{lmn} (d_{R, k} \Sigma_{li} \Sigma_{mj} d_{L,n}^\dagger +d_{L, k} \Sigma^\dagger_{li} \Sigma^\dagger_{mj} d_{R,n}^\dagger).
\end{align}
From the Lagrangian, we obtain the width of $qq(0^-) \to qq(0^+) \eta$ decay as
\begin{equation}
	\Gamma_{qq} = \frac{1}{6\pi} \left( \frac{m_1^2 +m_2^2}{f}\right)^2 \frac{|\bm{p}|}{M(0^-)^2}, \label{eq:wid_di_para}
\end{equation}
where $\bm{p}$ is the momentum of $\eta$, and $M(0^-)$ is the mass of diquark with spin-parity $0^-$.
The GT relation of diquark in the chiral limit is
\begin{equation}
	\frac{m_1^2 +m_2^2}{f} = \frac{[M(0^-)]^2 -[M(0^+)]^2}{2 f},
\end{equation}
where $M(0^+)$ is the mass of diquark with spin-parity $0^+$.
Using this relaton, Eq.~(\ref{eq:wid_di_para}) is rewritten as
\begin{equation}
	\Gamma_{qq} = \frac{2}{3 \pi} \left( \frac{M(0^-) -M(0^+)}{2f}\right)^2 \left( \frac{M(0^-) +M(0^+)}{2 M(0^-)} \right)^2 |\bm{p}|. \label{eq:width_qq}
\end{equation}
When $\langle \Sigma_{ij} \rangle = f \delta_{ij}$ is small compared with $M (0^+)$,
we can take $|\bm{p}| \to M(0^-) -M(0^+) \equiv \Delta M_{qq}$ and $M(0^-) + M(0^+) \to 2 M (0^-)$.
Then, the decay width is expressed by the mass difference $\Delta M_{qq}$ as
\begin{equation}
	\Gamma_{qq} = \frac{[\Delta M_{qq}]^3}{6\pi f^2} \label{eq:wid_diquark}.
\end{equation}
On the other hand, the width of $\Lambda_Q (1/2^-) \to \Lambda_Q (1/2^+) \eta$ decay is obtained from the Lagrangian of SHBs (\ref{eq:lag}) as
\begin{equation}
	\Gamma_{qqQ} = \frac{2}{3 \pi} \left( \frac{\Delta M}{2f}\right)^2 \frac{M(1/2^+)}{M(1/2^-)} |\bm{p}|, \label{eq:width_qqQ}
\end{equation}
where we used the GT relation in the chiral limit (\ref{eq:gtcl}).
In the heavy-quark limit, we can replace $|\bm{p}| \to \Delta M$ and $M(1/2^+) / M(1/2^-) \to 1$,
and obtain
\begin{equation}
	\Gamma_{qqQ} = \frac{[\Delta M]^3}{6 \pi f^2} \label{eq:wid_SHB}.
\end{equation}
Now, we can easily see a coincidence between Eqs.~(\ref{eq:wid_diquark}) and (\ref{eq:wid_SHB}) with an approximation $\Delta M_{qq} \simeq \Delta M$.

Next, let us include the flavor-symmetry breaking by $A > 1$.
The coupling constant of $ud (0^-) \to ud (0^+) \eta$ decay is obtained as
\begin{align}
	g_{ud} = \frac{m_1^2 +m_2^2}{f} &= \frac{\Delta [M_{1, 2}]^2 +\Delta [M_3]^2}{2 f (A +1)} \notag \\
	&=\frac{\Delta [M_3]^2}{2 f} \frac{\frac{\Delta [M_{1, 2}]^2}{\Delta [M_3]^2} +1}{A +1}, \label{eq:cc_ud}
\end{align}
where $\Delta [M_i]^2 = [M_i (0^-)]^2 -[M_i (0^+)]^2$ ($i = 1, \, 2, \, 3$).
From the inverse mass ordering $\Delta [M_{1, 2}]^2 < \Delta [M_3]^2$ and $A >1$, this coupling constant is smaller than the one expected from the GT relation of diquark: 
\begin{equation}
	\bar{g}_{ud} = \frac{m_2^2}{f} = \frac{\Delta [M_3]^2}{2f}. \label{eq:cc_ud_wo}
\end{equation}
Therefore, we expect that the effect of anomaly suppresses the decay width of $ud (0^-) \to ud (0^+) \eta$, similarly to that of $\Lambda_Q (1/2^-) \to \Lambda_Q (1/2^+) \eta$.

Adopting the prescription of $\eta$-$\eta^\prime$ mixing in Sec.~\ref{sec:eta}, the coupling constant of $ud (0^-) \to ud (0^+) \eta$ including the effect of mixing is obtained as
\begin{equation}
	g^{\rm{phys}}_{ud} = \frac{\xi_1 m_1^2 +\xi_2 m_2^2}{f}. \label{eq:cc_ud_phys}
\end{equation}
For the inverse mass ordering, $|m_1^2|>|m_2^2|$, $m_1^2>0$, and $m_2^2<0$ are hold~\cite{Harada:2019udr}.
Because $\xi_1 < 1$ and $\xi_2 > 1$ as shown in Sec.~\ref{sec:eta}, the effect of $\eta$-$\eta^\prime$ mixing also suppresses the decay width compared with the one calculated from the coupling constant naively expected from the GT relaion as
\begin{equation}
\bar{g}_{ud}^{\rm{phys}} = \frac{\xi_2 m_2^2}{f} = \xi_2 \bar{g}_{ud}. \label{eq:cc_ud_wo_phys}
\end{equation}

\begin{figure}[H]
  \begin{center}
   \includegraphics[bb=0 0 432 288, width=70mm]{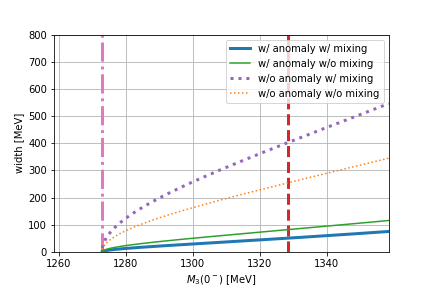}
  \end{center}
  \caption{Dependence of $ud (1/2^-) \to ud (1/2^+) \eta$ decay
  width on $M_3 (0^-)$. The curves correspond to those in Fig.~\ref{fig:charm_eta}.}
  \label{fig:diquark}
\end{figure}

We numerically evaluate the decay width of $ud (0^-) \to ud (0^+) \eta$ with taking $M_{1,2}(0^+) = 906$ MeV and $M_3 (0^+) = 725$ MeV~\cite{Harada:2019udr, Bi:2015ifa}.
We show the dependence of $ud (0^-) \to ud (0^+) \eta$ width on $M_3 (0^-)$ in Fig.~\ref{fig:diquark}.
The vertical thick-dashed red line is at $M_3 (0^-) = 1329$ MeV estimated from $M (\Lambda_c (1/2^-)) = 2890$ MeV,
while the vertical thick-dash-dotted magenta line is at the threshold of $ud (0^-) \to ud (0^+) \eta$ decay.
The thin-solid green and thin-dotted orange curves are plotted without $\eta$-$\eta^\prime$ mixing.
The thin-solid green curve is drawn by using the coupling constant in Eq.~(\ref{eq:cc_ud}) which includes the effect of anomaly, while the thin-dotted orange curve is by the one in Eq.~(\ref{eq:cc_ud_wo}) without anomaly.
The thick-solid blue and thick-dotted purple curves include the effect of $\eta$-$\eta^\prime$ mixing.
The thick-solid blue curve is drawn by using the coupling constant in Eq.~(\ref{eq:cc_ud_phys}) which includes the effect of anomaly, while the thick-dotted purple curve is by the one in Eq.~(\ref{eq:cc_ud_wo_phys}).
From Fig.~\ref{fig:diquark}, we see the suppression by the effects of anomaly and $\eta$-$\eta^\prime$ mixing similarly to the SHBs shown in Figs.~\ref{fig:charm_eta} and \ref{fig:bottom_eta}.
Because $\left[ (M (0^-) +M(0^+))/2 M (0^-) \right]^2$ in Eq.~(\ref{eq:width_qq}) is smaller than $M (1/2^+) / M (1/2^-)$ in Eq.~(\ref{eq:width_qqQ}), the widths of the diquark shown by the thin-dotted orange and thick-dotted purple curves are smaller than those of the SHBs in Figs.~\ref{fig:charm_eta} and \ref{fig:bottom_eta}.

\end{document}